\documentclass[crystals,review,accept,moreauthors,pdftex,10pt,a4paper]{mdpi} 
\firstpage{1} 

\articlenumber{x}
\doinum{10.3390/------}
\pubvolume{xx}
\pubyear{2016}
\copyrightyear{2016}
\externaleditor{Academic Editor: name}
\history{Received: date; Accepted: date; Published: date}

 \theoremstyle{mdpi}
 \newcounter{thm}
 \newcounter{ex}
 \newcounter{re}
 \theoremstyle{mdpidefinition}

\Title{Topology, Holonomy, and Quantum Walks}

\nolinenumbers

\Author{Graciana Puentes $^{1,*}$}

\address{%
$^{1}$ \quad Departamento de Fisica, Facultad de Ciencias Exactas y
Naturales, Pabellon 1, Ciudad Universitaria, 1428 Buenos Aires}
\corres{Correspondence: gracianapuentes@gmail.com}

\abstract{We present a review on the progress in the understanding and characterization of  holonomy and topology of a discrete-time quantum walk architecture, consisting of a unitary step given by a sequence of two non-commuting rotations in parameter space \cite{Puentesarxiv}. Unlike other similar systems recently studied in detail in the literature, this system does not present continous 1D topological boundaries, it only presents a discrete number of Dirac points where the quasi- energy gap closes. At these discrete points the topological winding number is not defined. Therefore, such discrete  points represent topological boundaries of dimension zero, and they endow the system with a non-trivial topology. We illustrate the non-trivial character of the system by calculating the Zak phase.  We also  propose a suitable experimental scheme to implement these ideas, and we present preliminary experimental data. }


\usepackage{lineno} 
\usepackage{graphicx,color}
\usepackage{color}   
\usepackage{hyperref}  

\definecolor{mygreen}{rgb}{0,0.5,0} 
\definecolor{myblue}{rgb}{0,0,1}
\definecolor{myred}{rgb}{1,0,0} 
\definecolor{mymagenta}{cmyk}{0,1,0,0.12}

\def\ba{\begin{eqnarray}}
\def\ea{\end{eqnarray}}

\def\lb{\label}
\def\be{\begin{equation}}
\def\ee{\end{equation}}
\def\thesection{\arabic{section}.}
\def\thesubsection{\arabic{section}.\arabic{subsection}}

\begin{document}

\section{Introduction}

Quantum physics attaches a phase to particles due to the complex nature of the Hilbert space. Phases arising during the quantum evolution of a particle can have different origins. A type of geometric phase, the so-called Berry phase \cite{Berry}, can be ascribed to quantum particles which return adiabatically to their initial state, but remember the path they took by storing this information on a geometric phase ($\Phi$), defined as \cite{Berry, Hannay}:
\begin{equation}
e^{i\Phi}=\langle \psi_{\mathrm{ini}}|\psi_{\mathrm{final}} \rangle.
 \end{equation}

 Geometric phases carry a number of implications: they modify material properties of solids, such as conductivity in graphene \cite{Berrygraphene}, they are responsible for the emergence of surface edge-states in topological insulators, whose surface electrons experience a geometric phase \cite{Berrytopoinsul}, they can modify the outcome of molecular chemical reactions  \cite{Berrychemestry}, and  could even have implications for quantum information technology, via the  Majorana particle \cite{Berrymayorana}, or can bear close analogies to gauge field theories and differential geometry \cite{BerryGauge}.\\

In this paper, we present a review on the progress in the understanding and characterization of dynamical effects, geometry, holonomy, and topology of a discrete-time quantum walk architecture, consisting of a unitary step given by a sequence of two non-commuting rotations in parameter space \cite{Puentesarxiv}. As pointed out in Ref \cite{Puentesarxiv}. page 2, this system does not present continous 1D topological boundaries. Unlike other systems recently studied in detail in the literature such as the "split-step" quantum walk  \cite{Kitagawa2,Kitagawa}, the system we report only presents a discrete number of Dirac points where the quasi-energy gap closes. Due to this apparent simplicity, the authors  concluded that the system was topologically trivial and a retraction was issued.\\

 Neverthereless, at these discrete Dirac points the topological winding number is not defined, therefore these discrete points represent topological boundaries of dimension zero. Since the system has toplogical boundaries it is topologically non-trivial. We demonstrate the non-trivial geometric landscape of the system by calculating different holonomic quantities such as the Zak phase \cite{Zak}. We propose a suitable experimental scheme for the implementation of the proposed ideas, and preliminary experimental data

\section{Holonomy, Topology and the Berry phase}
The concept of Berry and Zak phases are related to the mathematical concept of the holonomy of a  manifold. In the present section this important link is briefly described.
\\
 
\textit{The holonomy from a geometrical point of view:} As it is well known in differential geometry, the holonomy group $H_x$ at a point $x$ for an arbitrary oriented $n$-dimensional manifold $M$ endowed with metric $g_{ij}$
is defined by considering the parallel transport of a arbitrary vector field $V\in TM_x$ along all the possible closed curves $C$ starting and ending at $x$. The condition of parallel transport
is expressed as
\begin{equation}\label{pp}
t^\mu\nabla_\mu V=0,
\end{equation}
with $t^\mu$ the unit tangent vector of the curve $C$ and  $\nabla_\mu$ the Levi-Civita connection of ($M$, $g_{ij}$), i.e the unique connection which is torsion free and satisfies $\nabla g_{\mu\nu}=0$. By the Levi-Civita conditions together with (\ref{pp}) it follows directly that
$$
t^\mu \nabla_\mu(g_{ij}V^i V^j)=0.
$$
which is the statement that the norm of the vector field $||V||=g_{ij}V^i V^j$ is conserved during the travel along $C$. However, after the travel ends, the resulting vector $V_C$  will not coincide with $V$, but it will be rotated as $$V_C=R_x(C) V,$$ 

with $R(C)$ an element
of $SO(n)$. Thus, there is an assignment of a rotation matrix $R_x(C)$ corresponding to any pair $(x, C)$ with $C$ an arbitrary curve in the manifold. The possible rotations 
$R_x(C)$ at a fixed point $x$ obtained by considering all the possibles curves $C$ can be shown to be a group, which in fact  is equal or smaller than $SO(n)$ . This is known as  the holonomy group $H_x$ at $x$. If the manifold $M$ is simple connected, the holonomy groups at 
different points $x$ and $y$ are isomorphic and one simply speaks about the holonomy $H$ of $M$. Otherwise the definition of holonomy may be point dependent.

The concept of holonomy is more simply visualized for a manifold $M$ which is embedded in $R^n$. An example of this is the sphere $S^2$ with its canonical metric
\begin{equation}\label{se}
g_{S^2}=d\theta^2+\sin^2\theta d\phi^2.
\end{equation}
This sphere can be though as a surface $x_1^2+x_2^2+x_3^2=1$ embedded in $R^3$ and the canonical metric $g_{S^2}$ is simply the distance element of this surface. The holonomy element $H(C)$ for a given curve $C$ is intuitively a rotation $R(\alpha)$ with $\alpha(C)$ the solid angle subtended by $C$ at the center of the sphere. It is instructive to check this explicitly. For this, consider the unit vector  $r$ in $R^3$ parameterizing the points of the sphere
 $$
 r=(\sin\theta \cos\phi, \sin\theta \sin\phi, \cos\theta).
 $$
 Take a vector $V\in TM_x$ which will be parallel transported along a curve $C$ in $S^2$. If one describes the path by a parameter $t$, which can be though as the traveling "time" along the path, the the parallel transport condition can be expressed as follows. This vector should always be orthogonal to $r$, that is, $V\cdot r=0$, otherwise it will have a component orthogonal to the surface. However,  this is not enough since the vector $V$ may stay in the tangent plane $T_x S^2$ of any point $x$ visited during the travel, but a velocity $\Omega$ with a non zero component in the $r$ direction would make the vector rotate inside the planes. In order to avoid these rotations one should insure that $\Omega$  has no components in the direction of $r$, that is, $\Omega\cdot r=0$.  The total angular velocity $\Omega$ should not be zero however, since intuitively the vector $V$ has a component in $\dot{r}$ which is conserved, an $\dot{r}$ is making a rotation in $R^3$ and so does $V$. The two conditions stated above 
\begin{equation}\label{poro}
 \dot{V}=\Omega\times V,\qquad  \Omega=r\times \dot{r}.
 \end{equation}
For further applications in quantum mechanics, it is convenient to express 
the condition above in terms of the complex unit vector  $\psi$ defined by
$$
\psi=\frac{1}{2}(V+iV'),\qquad V'=r\times V.
$$
Then the  condition of parallel transport on the sphere (\ref{poro}) may be expressed alternatively as
\begin{equation}\label{par}
\Im(\psi^\ast \cdot d\psi)=0,\qquad \longrightarrow\qquad \Im(\psi^\ast \cdot d\psi)=0.
\end{equation}
Now, in order to find $\alpha(C)$ one may define a local orthogonal basis $u$ and $v$. For instance $u$ may lie on the parallel of latitude $\theta$ and $v$ the meridian of longitude $\phi$. These vectors are
explicitly given as
\be
u(r)=(-\sin\phi, \cos\phi, 0), 
\ee
\be
v(r)=(-\cos\theta \cos\phi,-\cos\theta \sin\phi, \sin \theta).
\ee
In these terms the phase $\alpha$ of $\psi$ may be defined as
$$
\psi=n \exp(i\alpha),\qquad n=\frac{1}{2}(u+iv).
$$
Note that the phase $\alpha$ may depend on the choice of $u$ and $v$, but the phase \emph{change} due to the transport along $C$ \emph{does not}. This phase change is given by
$$
\alpha(C)=\oint_C d\alpha=\Im(\oint n^\ast \cdot dn)=\Im\int \int_{Int(C)}dn\wedge dn^\ast.
$$
where in the last step the Stokes theorem was taken into account. Note that the last integrand is invariant under the gauge transformations
$$
n'=n \exp(i \mu(r)).
$$
By writing explicitly this integral in terms of our coordinate system it is obtained that
\be
\alpha(C)=\Im\int \int_{Int(C)}d\theta d\phi (\partial_\theta n^\ast \cdot\partial_\phi n-\partial_\theta n\cdot\partial_\phi n^\ast),
\ee
\be
\alpha(C)=\int\int_{Int C}\sin\theta d\theta d\phi.
\ee
which is clearly the solid angle subtended by $C$, as anticipated.
\\

\textit{Some uses of holonomy in quantum mechanics:} The notion of holonomy  can be generalized for more general connections $\widetilde{\nabla}$ than the Levi-Civita. These generalization may suited for dealing for quantum mechanical problems.  In the quantum mechanics set up one may replace the complex vector $\psi(\theta, \phi)$ by a state vector 
$|\psi(X)>$ where $X$ are the coordinates describing the parameter space of the problem. For instance, it the wave function $|\psi(X)>$ is describing the motion of electrons in the field of heavy ions, which can be consider effectively static, then the parameter $X$ is the position of these ions.

By introducing a complex basis $|n(X)>$ for any $X$,  one may define the relative phase of $|\psi(X)>$ 
$$
|\psi\rangle=|n(X)\rangle \exp(i\gamma).
$$
As before this phase is base dependent, but the holonomy to be defined below is instead independent on that choice. This holonomy is defined by an adiabatic travel around a circuit $C$ in the parameter space. After the travel,  the resulting wave function will acquire an additional phase due to the non trivial holonomy of such space. In other words
 $$
\langle \psi_{\mathrm{ini}}|\psi_{\mathrm{final}} \rangle=\exp(i\alpha(C)).
$$
The phase $\alpha(C)$ is known as the Berry phase \cite{Berry}. The condition of parallel transport (\ref{par}) becomes in this context
$$
\Im\langle\psi|d\psi\rangle=0.
$$
By simple generalization of the arguments given above in the differential geometrical context,  it follows that this phase is simply
\begin{equation}\label{V}
\alpha(C)=\int \int_{Int(C)}V, \qquad V=\Im<dn| \wedge |dn>.
\end{equation}
Thus, this phase is dependent on the path $C$, even with the adiabatic condition taken into account.

It may seem that the holonomy just defined was obtained without mentioning any metric tensor $g_{ij}$ in the Hilbert space in consideration. As we will discuss
below the natural language for such holonomy is in terms of principal bundles. However there exist a natural metric $g_{ij}$ in the parameter space of the problem. This issue was studied in \cite{Provost}  where the authors considered the following tensor
$$
T_{ij}=<\partial_i n|(1-|n><n|)|\partial_j n>=g_{ij}+i\frac{V_{ij}}{2}.
$$
This tensor is gauge invariant, that is, invariant under the transformation $$|n(X)>\to |n(X)>\exp(-i\mu(r)).$$
The real part of $T_{ij}$ is the form $V_{ij}$ whose flux gives the phase $\gamma(C)$. The imaginary part is a candidate for a metric tensor $g_{ij}$ \cite{Provost}.
In fact, one may define a "distance" between two states by the relation
$$
\Delta_{12}=1-|<\psi_1|\psi_2>|^2.
$$
The interpretation of this distance is intuitive. If two states $|\psi_1>$ and $|\psi_2>$ are identified when they differ by a global phase, and in this case $\Delta_{12}=0$.
In the case in which the overlap is minimal, the distance is maximal. Taking the limit $1\to 2$ and using the fact that the states are normalized gives that
\begin{equation}\label{er}
ds^2=<dn|(1-|n><n|)|dn>=T_{ij}dX^i dX^j=g_{ij}dX^i dX^j,
 \end{equation}
the last equality follows from the fact that the product of a symmetric tensor  such as $dX^i dX^j$ by an antisymmetric one such as $V_{ij}$ is zero.
It is interesting to note that for a two state spin system
$$
|+>=
\left(
\begin{array}{ccc}
  \cos\frac{\theta}{2} e^{i\frac{\phi}{2}}  \\
  \sin\frac{\theta}{2}e^{-i\frac{\phi}{2}}
\end{array}
\right),\qquad 
|->=
\left(
\begin{array}{ccc}
  \sin\frac{\theta}{2} e^{i\frac{\phi}{2}}  \\
  -\cos\frac{\theta}{2}e^{-i\frac{\phi}{2}}
\end{array}
\right)
$$
reduces to the definition (\ref{er}) gives the canonical metric on $S^2$ (\ref{se}), which is a nice consistency test. More general geometries are considered in \cite{Gibbons1}-\cite{Gibbons2}.
\\

There is an important subtle detail to be discussed, which is related to the use of the terms topology and geometry. The holonomy for a manifold $M$ in a global geometry context is geometrical, since the notion of parallel transport described above is related to the standard Levi-Civita connection $\nabla_i$, which is constructed directly in terms of the particular metric tensor $g_{ij}$ defined on $M$. However, such holonomy is not a topological invariant for $M$. In fact, there may exist two different complete metrics $g_{ij}$ and $g'_{ij}$ defined on the same manifold $M$ and possessing different holonomy groups. For instance any algebraic surface X which is compact, Kahler and with first real Chern class equal to zero admits a Ricci flat metric with holonomy $SU(3)$, due to the famous Yau theorem. The canonical metric on these surfaces are known, but none of them has holonomy $SU(3)$. In fact no compact metric with holonomy $SU(3)$ is known explicitly, since such metrics do not admit globally defined Killing vectors and thus they are highly non trivial. \\

In the quantum mechanical context however, the Berry phase may describe a topological phenomena in the following sense. The  description of the Berry phase made above bears a formal analogy with the concept of holonomy.  Nevertheless,  the precise notion for describing such phase is as the holonomy for an abstract connection in a principal bundle $P(U(1), X)$ with $X$ the parameter space $X$ of the problem. The curvature of this connection is the quantity $V$ defined in (\ref{V}). This quantity is known as the Berry curvature, it is gauge invariant and its flux is the Berry phase $\alpha(C)$. It turns out that for closed manifolds, these fluxes describe a Chern class of the bundle. These classes are invariant under gauge transformations and take integer values. Such classes describe inequivalent bundles over the parameter space $X$ and are topological, that is, they do not depend on the choice of the metric in the underlying manifold $X$. To describe these classes is beyond the scope of this work, extensive information can be found in \cite{Nakahara}, \cite{Berry}, in particular in connection with the physic of Dirac monopoles. We turn now to some applications of these concepts to real quantum mechanical problems.

\section{Discrete-time quantum walks}

~Discrete-time quantum walks (DTQWs) \cite{Aharonov} offer a versatile platform for the exploration of a wide range of non-trivial geometric and topological phenomena (experiment) \cite{Kitagawa} \cite{Crespi} \cite{Alberti}, and (theory) \cite{Kitagawa2,Obuse,Shikano2,Wojcik,MoulierasJPB,Grunbaum1,Grunbaum2}. Further, QWs are robust platforms for modeling a variety of dynamical processes from excitation transfer in spin chains \cite{Bose,Christandl} to energy transport in biological complexes \cite{Plenio}. They enable to study multi-path quantum inteference phenomena \cite{bosonsampling1,bosonsampling2,bosonsampling3,bosonsampling4}, and can provide for a route to validation of quantum complexity \cite{validation1,validation2}, and universal quantum computing \cite{Childs}. Moreover, multi-particle QWs warrant a powerful tool for encoding information in an exponentially larger space \cite{PuentesPRA}, and for quantum simulations in biological, chemical and physical systems \cite{PuentesOL}, in 1D and 2D geometries \cite{Peruzzo} \cite{Silberhorn2D}. \\

In this paper, we present a simple theoretical scheme for generation and detection of a non-trivial invariant geometric phase structure in 1D  DTQW architectures. The basic step in the standard DTQW is given by a unitary evolution operator $U(\theta)=TR_{\vec{n}}(\theta)$, where $R_{\vec{n}}(\theta)$ is a rotation along an arbitrary direction $\vec{n}=(n_{x},n_{y},n_{z})$, given by $$R_{\vec{n}}(\theta)=
\left( {\begin{array}{cc}
 \cos(\theta)-in_{z}\sin(\theta) & (in_{x}-n_{y})\sin(\theta)  \\
 (in_{x}+n_{y})\sin(\theta) & \cos(\theta) +in_{z}\sin(\theta)  \\
 \end{array} } \right), $$  in the Pauli basis \cite{Pauli}. In this basis, the y-rotation is defined by a coin operator of the form  \cite{Pauli}.
$$R_{y}(\theta)=
\left( {\begin{array}{cc}
 \cos(\theta) & -\sin(\theta)  \\
 \sin(\theta) & \cos(\theta)  \\
 \end{array} } \right). $$  This is  
followed by a spin- or polarization-dependent translation $T$ given by 
$$
T=\sum_{x}|x+1\rangle\langle x | \otimes|H\rangle \langle H| +|x-1\rangle \langle x| \otimes |V\rangle \langle V|,
$$
 where $H=(1,0)^{T}$ and $V=(0,1)^{T}$.
The evolution operator for a discrete-time step is equivalent to that generated by a Hamiltonian $H(\theta)$, such that $U(\theta)=e^{-iH(\theta)}$ ($\hbar=1$), with $$H(\theta)=\int_{-\pi}^{\pi} dk[E_{\theta}(k)\vec{n}(k).\vec{\sigma}] \otimes |k \rangle \langle k|$$ and $\vec{\sigma}$ the Pauli matrices, which readily reveals the spin-orbit coupling mechanism in the system.~The quantum walk described by $U(\theta)$ has been realized experimentally in a number of systems \cite{photons,photons2,ions} \cite{coldatoms}, and has been shown to posses chiral symmetry, and display Dirac-like dispersion relation given by $\cos(E_{\theta}(k))=\cos(k)\cos(\theta)$. \\

\subsection{Split-step quantum walk}

We present two different examples of non-trivial geometrical phase structure in the holonomic sense, as described in the previous section. The first DTQW protocol consists of two consecutive spin-dependent translations $T$ and rotations $R$, such that the unitary step becomes $U(\theta_1,\theta_2)=TR(\theta_1)TR(\theta_2)$, as described in detail in \cite{Kitagawa2}. The so-called ``split-step" quantum walk, has been shown to possess a non-trivial topological landscape given by topological sectors which are delimited by continuous 1D topological boundaries. These topological sectors are characterized by topological invariants, such as the winding number, taking integer values $W=0,1$. The dispersion relation for the split-step quantum walk results in \cite{Kitagawa2}:
$$
 \cos(E_{\theta,\phi}(k))=\cos(k)\cos(\theta_1)\cos(\theta_2)-\sin(\theta_1)\sin(\theta_2).
$$
The 3D-norm for decomposing the quantum walk Hamiltonian of the system in terms of Pauli matrices $H_{\mathrm{QW}}=E(k)\vec{n} \cdot \vec{\sigma}$  becomes \cite{Kitagawa}: \\
\begin{equation}
\begin{array}{ccc}
n_{\theta_1,\theta_2}^{x}(k)&=&\frac{\sin(k)\sin(\theta_1)\cos(\theta_2)}{\sin(E_{\theta_1,\theta_2}(k))}\\
n_{\theta_1,\theta_2}^{y}(k)&=&\frac{\cos(k)\sin(\theta_1)\cos(\theta_2)+\sin(\theta_2)\cos(\theta_1)}{\sin(E_{\theta_1,\theta_2}(k))}\\
n_{\theta_1,\theta_2}^{z}(k)&=&\frac{-\sin(k)\cos(\theta_2)\cos(\theta_1)}{\sin(E_{\theta_1,\theta_2}(k))}.\\
\end{array}
 \end{equation}
The dispersion relation and topological landscape for the split-step quantum walk was analyzed in detail in \cite{Kitagawa2}. We now turn to our second example. \\

\subsection{Quantum walk with non-commuting rotations}

The second example consists of two consecutive non-commuting rotations in the unitary step of the DTQW. The second rotation along the x-direction by an angle $\phi$, such that  the unitarity step becomes $U(\theta,\phi)=TR_{x}(\phi)R_{y}(\theta)$, where $R_{x}(\phi)$ is given in the same basis \cite{Pauli} by:
$$R_{x}(\phi)=
\left( {\begin{array}{cc}
 \cos(\phi) & i\sin(\phi)  \\
i \sin(\phi) & \cos(\phi)  \\
 \end{array} } \right).$$ The modified dispersion relation becomes:
\begin{equation}
 \cos(E_{\theta,\phi}(k))=\cos(k)\cos(\theta)\cos(\phi)+\sin(k)\sin(\theta)\sin(\phi),
\end{equation}
 where we recover the Dirac-like dispersion relation for $\phi=0$, as expected.
~The 3D-norm for decomposing the Hamiltonian of the system in terms of Pauli matrices  becomes: 
\begin{equation}
\begin{array}{ccc}
n_{\theta,\phi}^{x}(k)&=&\frac{-\cos(k)\sin(\phi)\cos(\theta)+\sin(k)\sin(\theta)\cos(\phi)}{\sin(E_{\theta,\phi}(k))}\\
n_{\theta,\phi}^{y}(k)&=&\frac{\cos(k)\sin(\theta)\cos(\phi)+\sin(k)\sin(\phi)\cos(\theta)}{\sin(E_{\theta,\phi}(k))}\\
n_{\theta,\phi}^{z}(k)&=&\frac{-\sin(k)\cos(\theta)\cos(\phi)+\cos(k)\sin(\theta)\sin(\phi)}{\sin(E_{\theta,\phi}(k))}.\\
\end{array}
 \end{equation}

\begin{figure} 
\centering
\includegraphics[width=0.6\linewidth]{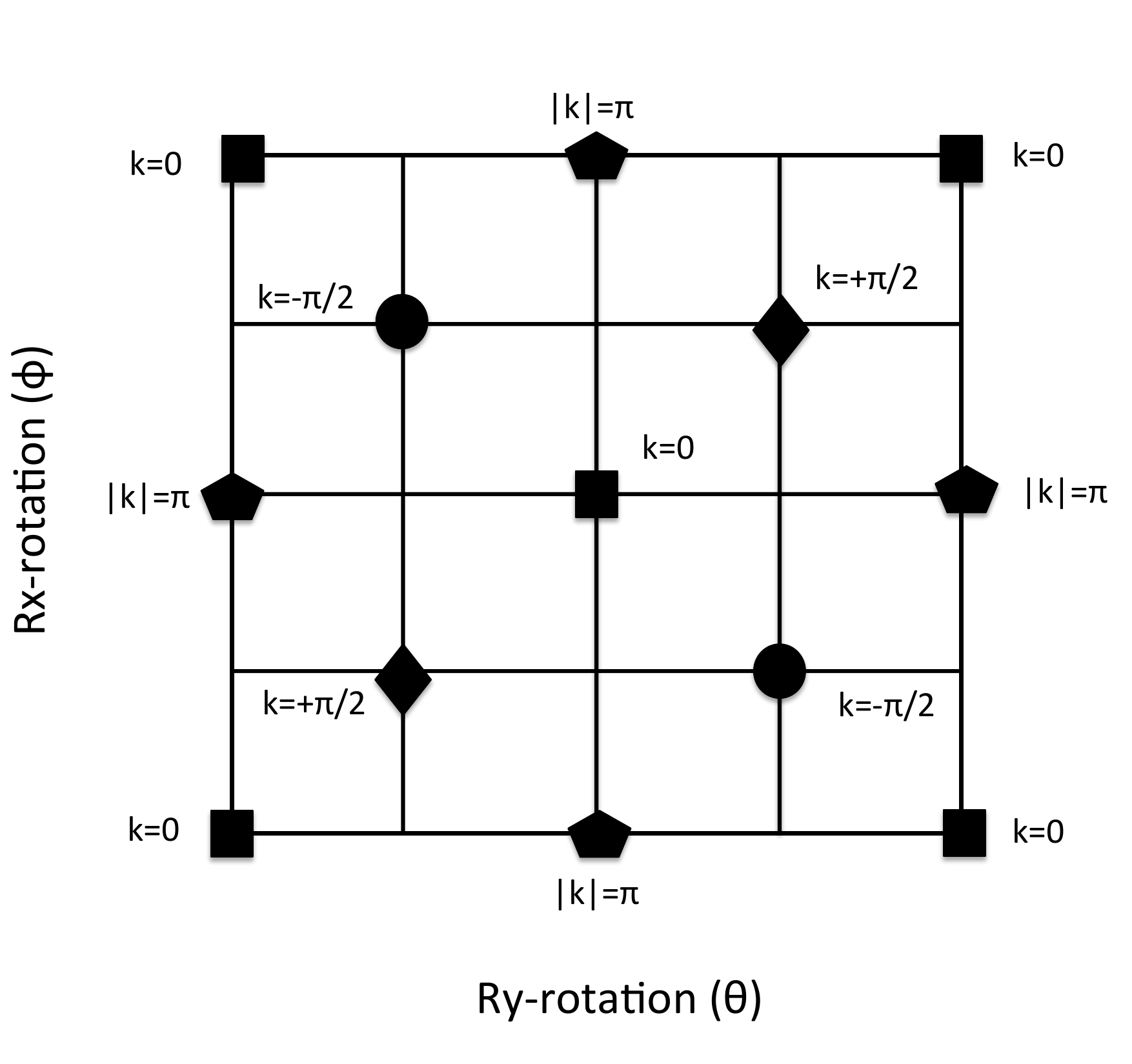} \caption{Non-trivial phase diagram for the quantum walk with consecutive non-commuting rotations, indicating gapless Dirac points where quasi-energy gap closses for different values of quasi-momentum: Squares ($k=0$), pentagons ($|k|=\pi$), romboids ($k=+\pi/2$), circles ($k=-\pi/2$). These discrete Dirac points represent topological boundaries of dimension zero. They endow the system with a non-trivial topology.  }
\end{figure}

As anticipated, this system has a non-trivial phase diagram with a larger number of gapless points for different momenta as compared to the system consisting of a single rotation. Each of these gapless points represent topological boundaries of dimension zero, where topological invariants are not defined. Unlike the "split-step" quantum walk described previously, this system does not contain continuous topological boundaries. We calculated analitically the gapless Dirac points and zero-dimension topological boundaries for the system. Using basic trigonometric considerations, it can be shown that thequasi-energy gap closes at 13 discrete points, for different values of quasi-momentum $k$. The phase diagram indicating the Dirac points where the gap closes for different momentum values is shown in Fig. 1. Squares correspond to Dirac points for $k=0$, circles correspond to Dirac points for $k=-\pi/2$, romboids correspond to Dirac points for $k=+\pi/2$, and pentagons correspond to Dirac points for $|k|=\pi$.  This geometric  structure in itself is novel and topologically non-trivial. Moreover, it has not been studied in detail before. \\

\section{Zak Phase Calculation}

We will now give expressions for the Zak Phase in two different scenarios. These scenarios are casted by the following hamiltonian
\begin{equation}
H\sim n_x \sigma_x+n_y \sigma_y+ n_z \sigma_z,
\end{equation}
The hamiltonian to be described differ by a multiplying factor and by the expression of the $n_i$.
But since the eigenvectors are the only quantities of interest for the present problem, the overall constants of this Hamiltonian can be safely ignored.
Now, our generic hamiltonian is given by the matrix
\begin{equation}
H=
\left(
\begin{array}{cc}
  n_z \qquad    n_x-in_y\\
 n_x+i n_y \qquad   -n_z   
\end{array}
\right),
\ee
and has the following eigenvalues
\begin{equation}
\lambda=\pm \sqrt{n_x^2+n_y^2+n_z^2}
\end{equation}
The normalized eigenvectors then result
\begin{equation}
|V_\pm>= 
\left(
\begin{array}{cc}
  \frac{n_x+i n_y}{\sqrt{2n_x^2+ 2n_y^2+2n_z^2\mp 2n_z\sqrt{n_x^2+n_y^2+n_z^2}}}    \\
  \frac{n_z\mp \sqrt{n_x^2+n_y^2+n_z^2}}{\sqrt{2n_x^2+ 2n_y^2+2n_z^2\mp 2n_z\sqrt{n_x^2+n_y^2+n_z^2}}}   
\end{array}
\right)
\end{equation}

Note that the scaling $n_i\to\lambda n_i$ does not affect the result, as should be. This follows from the fact that two hamiltonians
related by a constant have the same eigenvectors.

The Zak phase ($\Phi_{Zak}=Z$) for each band ($\pm$) can be expresssed as: 

\begin{equation}
Z_{\pm}=i\int_{-\pi/2}^{\pi/2} (<V_{\pm}|\partial_k V_{\pm}>) dk.
\end{equation}
We will now apply these concepts to some specific examples.

\subsection{Split-step Quantum Walk}

We first consider the split-step quantum walk. This corresponds to a quantum walk with unitary step give by $U(\theta_1, \theta_2)=TR(\theta_1)TR(\theta_2)$, as proposed in \cite{Kitagawa2}. In this example the normals $n_i$ are of the following form

\begin{equation}
\begin{array}{ccc}
n_{\theta_1,\theta_2}^{x}(k)&=&\frac{\sin(k)\sin(\theta_1)\cos(\theta_2)}{\sin(E_{\theta_1,\theta_2}(k))}\\
n_{\theta_1,\theta_2}^{y}(k)&=&\frac{\cos(k)\sin(\theta_1)\cos(\theta_2)+\sin(\theta_2)\cos(\theta_1)}{\sin(E_{\theta_1,\theta_2}(k))}\\
n_{\theta_1,\theta_2}^{z}(k)&=&\frac{-\sin(k)\cos(\theta_2)\cos(\theta_1)}{\sin(E_{\theta_1,\theta_2}(k))}.\\
\end{array}
 \end{equation}

We consider the particular case that $n_z=0$. By taking one of the angle parameters such that $n_z=0$, it follows that the eigenvectors of the Hamiltonian are:

\begin{equation}
|V_\pm>= 
\frac{1}{\sqrt{2}}\left(
\begin{array}{cc}
  e^{-i\phi(k)}   \\
  \mp 1 
\end{array}
\right),\qquad 
\tan\phi(k)=\frac{n_y}{n_x}.
\end{equation}
There are two choices for $n_z=0$, which are $\theta_1=0$ or $\theta_2=0$.
The Zak phase for each band takes the same value and results in:

\begin{equation}
Z=Z_{\pm}=i\int_{-\pi/2}^{\pi/2}dk <V_{\pm}|\partial_{k} V_{\pm}>
\end{equation}
\begin{equation}
Z=i\int_{-\pi/2}^{\pi/2} dk <V_{\pm}|\partial_k V_{\pm}>=\phi(-\pi/2)-\phi(\pi/2),
\end{equation}
from where it follows that
\begin{equation}
Z=\frac{\tan(\theta_2)}{\tan(\theta_1)}.
\end{equation}
A plot of the Zak phase is presented in Fig. 2 (a).

\subsection{Quantum walk with non-commuting rotations}

The unitary step as described in the introduction results in $U(\theta,\phi)=TR_{x}(\phi)R_{y}(\theta)$.  The norms $n_i$ are of the following form
\be
n_x=-\cos(k) a+\sin(k)b,
\ee
\be
 n_y=\cos(k) b+\sin(k) a,
\ee
\be
n_z=\cos(k) c-\sin(k) d,
\ee
with
\be
a=\sin(\phi)\cos(\theta)
\ee
\be\lb{ang}
 b=\cos(\phi)\sin(\theta),
 \ee
\be
 c=\sin(\phi)\sin(\theta),
\ee
 \be 
d=\cos(\phi)\cos(\theta).
 \ee

the  angular functions defined above. The numerator $N_1$ is given by
\begin{equation}\lb{n1}
N_1=n_x+in_y=- \exp(-i k)(a-i b),
\end{equation}
The remaining numerator $N_2$ is
$$
N_2=n_z\mp \sqrt{n_x^2+n_y^2+n_z^2}=\cos(k) c-\sin(k) d
$$
\be\lb{n2}
\mp \sqrt{a^2+b^2+c^2 \cos^2(k)+d^2 \sin^2(k)-\sin(2k)cd}
\ee
On the other hand, the denominator  $D$ is reduced to 
$$
D_{\pm}=\sqrt{2n_x^2+ 2n_y^2+2n_z^2\mp 2n_z\sqrt{n_x^2+n_y^2+n_z^2}}
$$
$$
=\bigg(a^2+b^2+c^2 \cos^2(k)+d^2 \sin^2(k)-\sin(2k)cd
$$
$$
\mp (\cos(k) c-\sin(k) d)
$$
\be\lb{d}
\times \sqrt{a^2+b^2+c^2 \cos^2(k)+d^2 \sin^2(k)-\sin(2k)cd}\bigg)^{\frac{1}{2}}.
\ee
By taking into account these expressions, one finds that the eigenvectors may be expressed simply as
\begin{equation}
|V_\pm>= 
\left(
\begin{array}{cc}
 \frac{N_1}{D_\pm}    \\
  \frac{N_2}{D_\pm} 
\end{array}
\right),\qquad 
<V_\pm|=\bigg(\frac{N^\ast_1}{D_\pm},\; \frac{N_2}{D_\pm}\bigg)
\end{equation}
Then the calculation of the Zak phase for each band results in:
$$
Z=Z_{\pm} =i\int_{-\pi/2}^{\pi/2}dk <V_{\pm}|\partial_k V_{\pm}>
$$

which requires to know the following quantities
$$
Z_\pm=i\int \bigg(\frac{N_1^\ast}{D_\pm^2}\partial_k N_1+\frac{N_2}{D_\pm^2}\partial_k N_2
$$
\begin{equation}
-\frac{(|N_1|^2+|N_2|^2)}{D_\pm^3}\partial_k D_\pm\bigg) dk,
\end{equation}
This expression can be simplified further. Due to  (\ref{n1}) it follows that the first term is real.  However, an inspection of (\ref{n2}) shows that the last two terms are purely imaginary. 
Since the overall phase should be real, it follows that these terms should cancel. This can be seen by taking into account that:
\be\lb{fb}
D_\pm=\sqrt{|N_1|^2+|N_2|^2},\qquad \partial_k |N_1|^2=0,
\ee
together with the fact that $N_2$ is real. Then  
\begin{equation}
\partial_k D_\pm=\frac{2N_2\partial_k N_2}{2D_\pm},
\end{equation}
where (\ref{fb}) has been taken into account. Therefore
$$
Z_\pm=i\int \bigg(\frac{N_1^\ast}{D_\pm^2}\partial_k N_1+\frac{N_2}{D_\pm^2}\partial_k N_2
$$
\begin{equation}
-\frac{(|N_1|^2+|N_2|^2)}{D_\pm^4}N_2\partial_k N_2\bigg) dk,
\end{equation}
but since  $D_\pm^2=|N_1|^2+|N_2|^2$ a simple calculation shows that the last two terms cancel each other. Thus the phase is
\begin{equation}
Z_\pm=i\int \frac{N_1^\ast\partial_k N_1}{D_\pm^2}dk.
\end{equation}
By taking into account  (\ref{d}) the Zak phase is expressed as
\begin{equation}
Z_\pm=\int \frac{|N_1|^2}{D_\pm ^2}dk=\int \frac{(a^2+b^2)dk}{D_\pm^2} 
\end{equation}
 
We note that in this example the case $n_{z}=0$ is completely different than in the previous case, as it returns a trivial Zak phase $Z=\pi$, since the k-dependence vanishes. We note that for this system the Zak phase landscape can be obtained by numerical integration. In particular, at the Dirac points indicated in Figure 1, the Zak phase is not defined.\\

\begin{figure} 
\label{fig:2}
\centering
\includegraphics[width=0.8\linewidth]{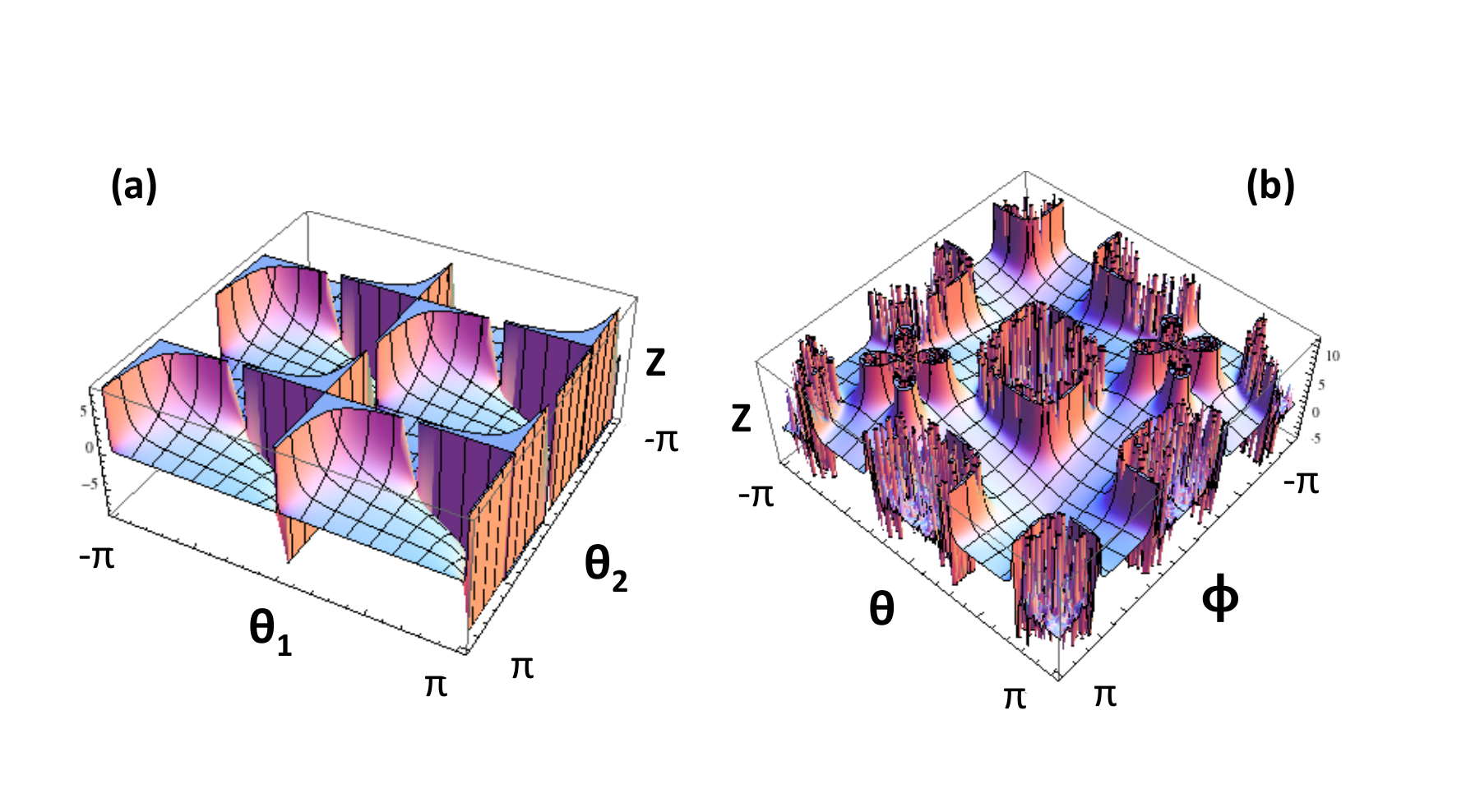} \caption{(a) Non-trivial geometric Zak phase landscape for "split-step" quantum walk, obtained analytically. (b) Non-trivial geometric Zak phase landscape for the quantum walk with non-commuting rotations, obtained by numeric integration.}
\end{figure}

\noindent A plot of the Zak phase $\Phi_{Zak}$ is shown in Fig. 2, for parameter values $\theta_{1,2}=[-\pi, \pi]$, and $\phi=[-\pi, \pi]$. (a) Zak phase for split-step quantum walk, given by the analytic expression $Z=\frac{\tan(\theta_2)}{\tan(\theta_1)}$, (b) Zak phase for quantum walk with non-commuting rotation obtained by numerical integration of  expression Eq. 4.41.\\

\subsection{Discussion}

It is well known that the Zak phase is not a geometric invariant, since it depends on the choice of origin of the Brillouin zone. However, a geometric invariant can be defined in terms of the Zak phase \emph{difference} between two states ($|\psi^{\bf{1}} \rangle, |\psi^{\bf{2}} \rangle$) which differ on a geometric phase only. Generically, the Zak phase difference between two such states can be written as $\langle  \psi^{\bf{1}} |\psi^{\bf{2}}  \rangle=e^{i|\Phi_{Zak}^{\bf{1}} - \Phi_{Zak}^{\bf{2}}|}$. We stress that by geometric invariance, we refer to properties that  do not depend on the choice of origin of the Brillouin zone but only on relative distances between points in the Brillouin zone.  \\

\noindent A simple experimental scheme to measure the Zak phase difference between states at any given step $N$ can be envisioned. For any choice of origin of the Brillouin zone, the system can be prepared by unitary evolution operators characterized by rotation parameters corresponding to either of the four adjacent Dirac points. A different geometric phase will be accumulated at each adjacent Dirac point. This phase difference can be measured by recombining the states, in the case of photons by interferring the states via a Mach-Zehnder interferometer. A suitable scheme for detection of the Zak phase difference in a photonic system is described in \cite{HolonomicWhite}. \\

\section{Previous Experimental Realizations}

The system here investigated, consisting of two non-commuting rotations $(R_{y}(\theta),R_{x}(\phi))$ in a discrete-time quantum walk for the study of geomtric phases in quantum walks was first proposed and studied in detail in a proposal originally introduced by Puentes \emph{et al.} in the year 2013 \cite{Puentesarxiv}. In the original contribution, the authors proposed and sucessfully realized a novel experimental scheme to implement the two consecutive non-commuting rotations in a photonic quantum walk for the first time \cite{Puentesarxiv} \\

\begin{figure} [h!]
\label{fig:1}
\centering
\includegraphics[width=0.6\linewidth]{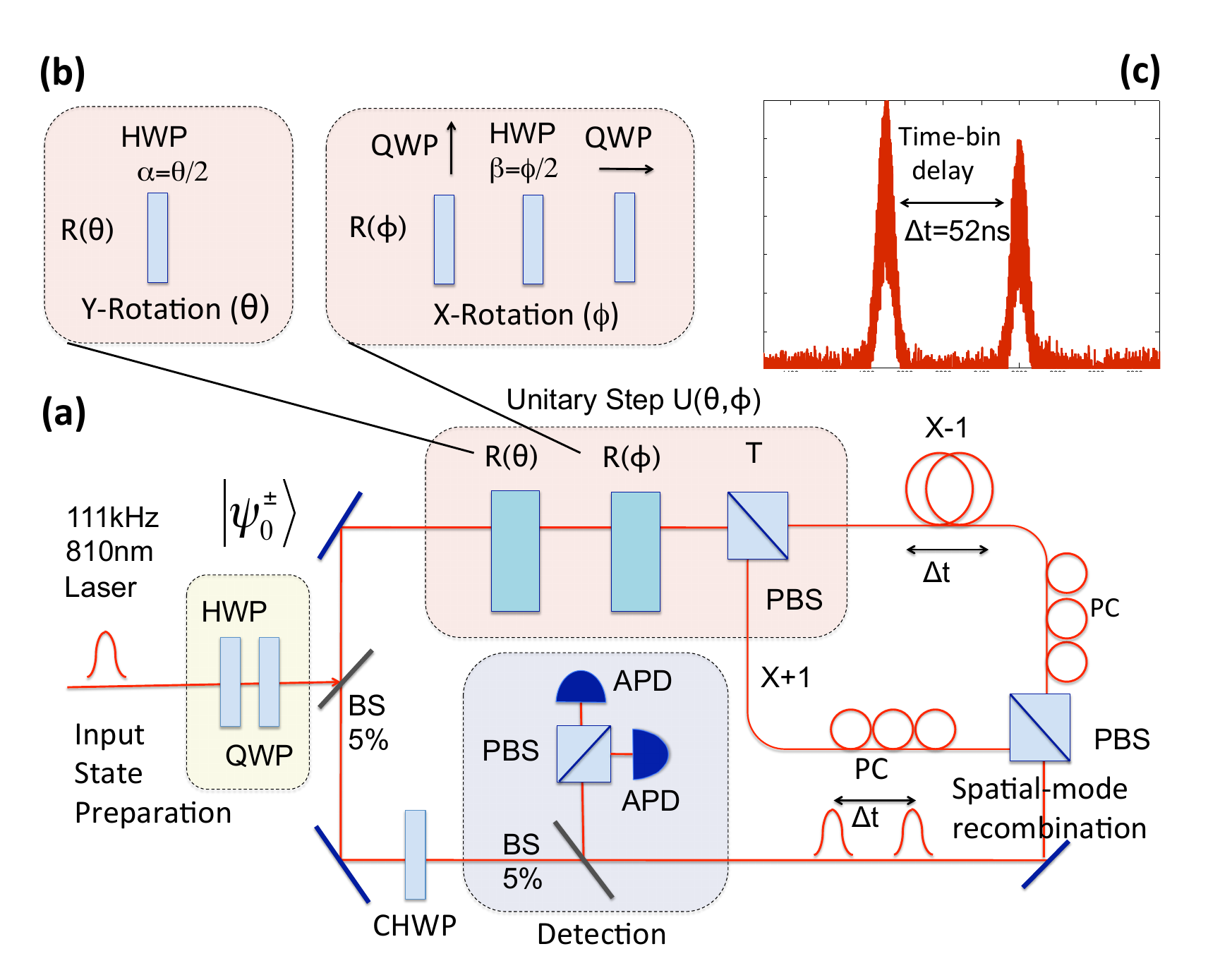} \caption{(a) Schematic of experimental setup \cite{Puentesarxiv6}. (b) Implementation of non-commuting rotations: $R_{y}(\theta)$ is implemented via a HWP at angle $\alpha=\theta/2$. $R_{x}(\phi)$ is implemented by a sequence of QWPs with fast axes oriented vertically and horizontally, respectively. In between the QWPs, a HPW oriented at $\beta=\phi/2$ determines the angle for the second rotation. (c) Histrogram of arrival times, after a trigger event at $t=0$.  }
\end{figure}

The experimental scheme implemented by the authors in Ref. \cite{Puentesarxiv}  is based on a time-multiplexed quantum walk realization introduced in Ref.  \cite{Silberhorn} (see Fig. 5 (a)). This scheme allows to implement a large number of steps in a compact architecture, thus improving upon previous realizations \cite{Kitagawa}.~Equivalent single-photon states are generated with an attenuated pulsed diode laser centered at 810 nm and with 111 kHz repetition rate (RR).  The initial state of the photons is controlled via half-wave plates (HWPs) and quarter-wave plates (QWPs),  to produce eigen-states of chirality $|\psi_{0}^{\pm}\rangle=|0\rangle \otimes 1/\sqrt{2}(|H\rangle \pm i|V\rangle)$. Inside the loop, the first rotation ($R_{y}(\theta)$) is implemented by a HWP with its optical axis oriented at an angle $\alpha=\theta/2$.~The rotation along the x-axis  ($R_{x}(\phi)$) is implemented by a combination of two QWPs with axes oriented horizontally(vertically), characterized by Jones matrices of the form 
$
\left( {\begin{array}{cc} 1 & 0  
\\ 0 & (-)i  \\ 
\end{array} } \right)$ (Fig. 5 (b)). In between the QWPs, a HWP oriented at $\beta=\phi/2$ determines the angle for the x-rotation. The spin-dependent translation is realized in the time domain via a polarizing beam splitter (PBS) and a fiber delay line, in which horizontally polarized light follows a longer path. The resulting temporal difference  between both polarization components corresponds to a step in the spatial domain  ($x \pm 1$). Polarization controllers (PC) are introduced to compensate for arbitrary polarization rotations in the fibers. After implementing the time-delay the time-bins are recombined in a single spatial mode by means of a second PBS and are re-routed into the fiber loops by means of silver mirrors. After a full evolution the photon wave-packet is distributed over several discrete positions, or time-bins. The detection is realized by coupling the photons out of the loop by a beam sampler (BS) with a probability of 5$\%$ per step. Compensation HWPs (CHWPs) are introduced to correct for dichroism at the beam samplers (BS).~We employ two avalanche photodiodes (APDs) to measure the photon arrival time and polarization properties. The probability that a photon undergoes a full round-trip is given by the overal coupling efficiency  ($>70 \%$) and the overall losses in the setup resulting in $\eta= 0.50$.  The average photon number per pulse is controlled via neutral density filters and is below $\langle n \rangle <0.003$ for the relevant iteration steps  ($N=7$) to ensure negligible contribution from multi-photon events. 

\begin{figure} [h!]
\label{fig:1}
\centering
\includegraphics[width=0.7\linewidth]{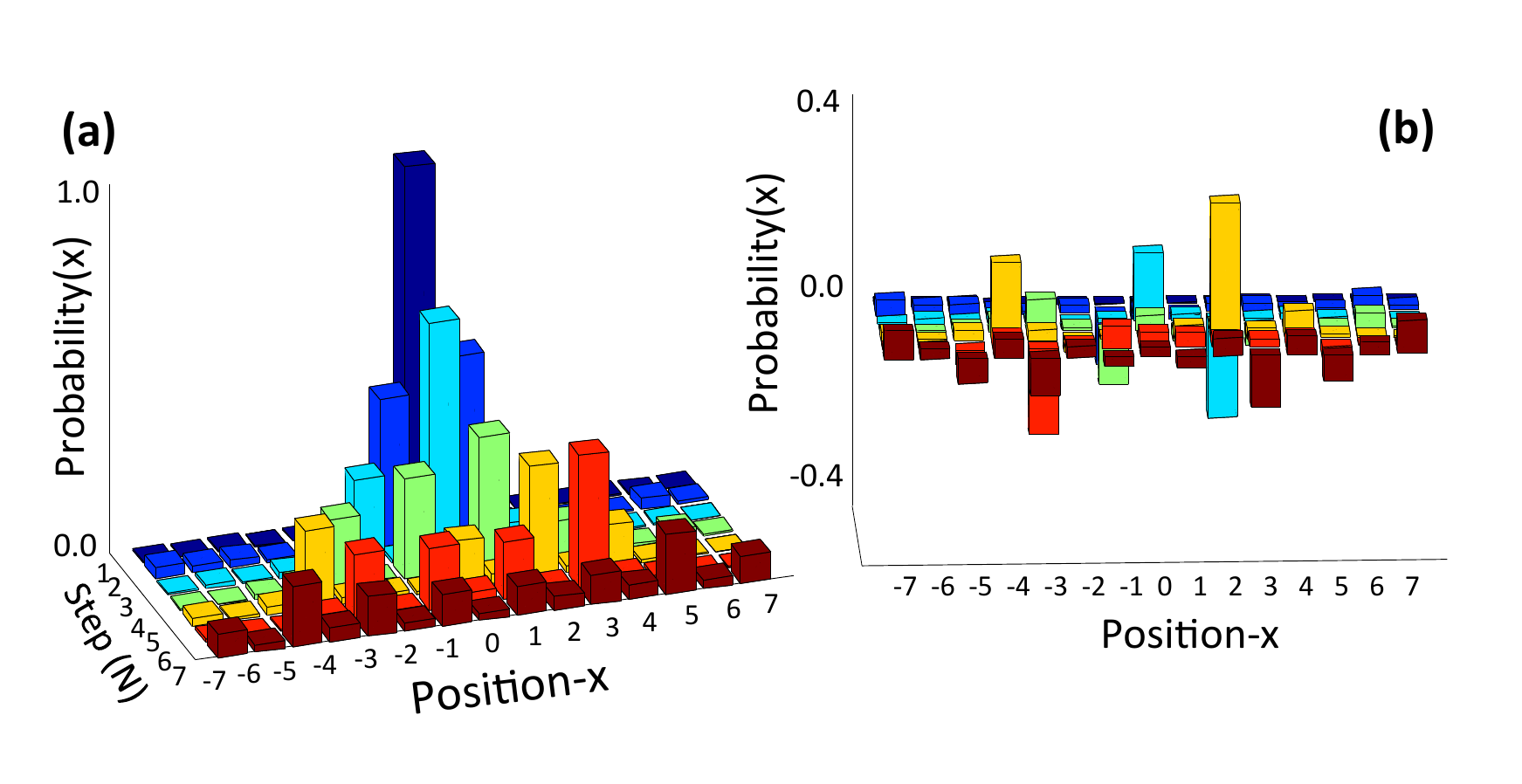} \caption{(a) Measured proability distributions for $N=7$ steps in  Hadamard QW  with $\theta=\pi/4, \phi=0$, and input state $|\psi_{0}^{+}\rangle$. (b) Difference between experiment and theory is within $20 \%$, and is mainly ascribed  to different soures of polarization dependent losses, spurious reflections, and shot-noise.}
\end{figure}

We characterized the round-trip time (RTT=$750$ ns) and the time-bin distance (TBD=$52$ ns) with a fast Oscilloscope (Lecroy 640ZI, 4GHz). The RTT, and the laser RR determine the maximum number of steps that can be observerd in our system ($N_{\mathrm{max}}=12$).  Figure 5 (c), shows typical time-bin traces obtained from time-delay histogram recorded  with 72 ps resolution. The actual number of counts was obtained by integrating over a narrow window. We first implemented the Hadamard quantum walk, by setting $\theta=\pi/4$ and $\phi=0$. This is shown in Fig. 6 (a) for the first $N=7$ steps with no numerical corrections for systematic errors, after background subtraction. We compare the theoretical and experimental probability distributions via the similariy $S=[\sum_{x} \sqrt{P_{\mathrm{theo}}(x)P_{\mathrm{exp}}(x)}]^2$, with $S=0(1)$ for orthogonal(identical) distributions \cite{Silberhorn2D}, typically obtaining $S  \approx 0.85$. The difference between raw data and theory are displayed in Fig. 6 (b). Experimental errors can be explained in terms of asymmetric coupling, imperfect polarization-rotation compensation in the fibers, unequal efficiency in the detectors, and other sources of polarization dependent losses, in addition to shot-noise. Uncontrolled reflections are a main source of error. We removed this by subtracting the counts of the two APDs, and filtering peaks located at positions different from the RTT and the TBD during data analysis.

\section{Conclusions}
We have reported in the sequential progress in the understanding of the topology and holonomy of a novel system consisting of a discrete-time quantum walk with consecutive non-commuting rotations. While we do not expect localization phemena in our system for the case of a stationary coin operation and  large number of steps \cite{BrunPRL2003}, we note that the system has a non-trivial topology due to the existence of topological boundaries of dimension zero, and we do predict the existence of geometric invariant structures. We argue that such invariants can be directly measured. We also propose a novel and roboust experimental scheme for the implementation of our proposal based on the original contributions in Ref. \cite{Puentesarxiv}.

\section{Acknowledgements}

The authors gratefully acknowledge O. Santillan, M. Saraceno and M. Hafezi. GP gratefully acknowledges financial support from PICT2014-1543 grant, PICT22015-0710 grant, UBACYT PDE 2015 award, and Raices programme. 


\section{References}
 
\begin{thebibliography}{11}

\bibitem{Puentesarxiv} Puentes \emph{et al.}, (Retracted article. See vol. 113, pg. 9901, 2014)  Phys. Rev. Lett. \textbf{112} 120502 (2014); arxiv/1311.7675 (2013). 

\bibitem{Berry} M. V. Berry, J. Phys. A \textbf{18} 15 (1985).

\bibitem{Hannay} J. Hannay, J. Phys. A \textbf{18} 221 (1985).

\bibitem{Provost} J. Provost and G. Vallee Comm. Math. Phys 76 (1980) 289.

\bibitem{Gibbons1} C. Bouchiat and G. Gibbons J. Phys. France 49 (1988) 187.

\bibitem{Gibbons2} D. Page Phys. Rev. A 36 (1987) 3479.

\bibitem{Berin} M .V. Berry  in "Geometric Phases in Physics" A. Shapere and F. Wilczek (Editors) World Scientific 1989.
 
\bibitem{Nakahara} M. Nakahara "Geometry, Topology and Physics" Graduate Student Series in Physics 1990.

\bibitem{Berrygraphene} Y. Zhang, Y.-W. Tan, H. L. Stormer, and P. Kim, Nature \textbf{438} 201 (2005).

\bibitem{Berrytopoinsul} C. L. Kane, and E. J. Mele, Phys. Rev. Lett. \textbf{95}, 146802 (2005); B. Bernevig \emph{et al.}, Science \textbf{314}, 1757 (2006); M. K\"{o}ning \emph{et al.}, Science \textbf{318}, 766 (2007).

\bibitem{Berrychemestry} G. Delacretaz, E. R. Grant, R.L. Whetten, L. W\"{o}ste, and J. W.  Zwanziger, Phys. Rev. Lett. \textbf{56},  2598 (1986).

\bibitem{Berrymayorana} S. Nadj-Perge  \emph{et al.}, Science \textbf{346}, 602 (2014).

\bibitem{BerryGauge} B. Simon, Phys. Rev. Lett. \textbf{51}, 2167 (1983).


\bibitem{Kitagawa} T. Kitagawa \emph{et al.}, Nature Communications \textbf{3}, 882 (2012). 


\bibitem{Zak} J. Zak, Phys. Rev. Lett. \textbf{62}, 2747 (1989).

\bibitem{Aharonov}Y. Aharonov, L. Davidovich, and N. Zagury, Phys. Rev. A \textbf{48}, 1687 (1993).


\bibitem{Crespi} A. Crespi \emph{et al.}, Nature Photon. \textbf{7}, 322 (2013).

\bibitem{Alberti} M. Genske \emph{et al.}, Phys. Rev. Lett. \textbf{110}, 190601 (2013).

\bibitem{Kitagawa2} T. Kitagawa, M. S. Rudner, E. Berg, and E. Demler, Phys. Rev. A \textbf{82}, 033429 (2010).

\bibitem{Obuse}H. Obuse and N. Kawakami, Phys. Rev. B \textbf{84}, 195139 (2011).

\bibitem{Shikano2} Y. Shikano, K. Chisaki, E. Segawa, and N. Konno, Phys. Rev. A \textbf{81}, 062129 (2010).

\bibitem{Asboth} J. K. Asb\'{o}th, Phys. Rev. B \textbf{86}, 195414 (2012).

\bibitem{Wojcik}A. W\'{o}jcik \emph{et al.}, Phys. Rev. A \textbf{85}, 012329 (2012).

\bibitem{MoulierasJPB}S. Moulieras, M. Lewenstein, and G. Puentes, J. Phys. B \textbf{46}, 104005 (2013).

\bibitem{Bose} S. Bose, Phys. Rev. Lett. \textbf{91}, 207901 (2003).

\bibitem{Christandl} M. Christandl, N. Datta, A. Ekert, and A. J. Landahl, Phys. Rev. Lett. \textbf{92}, 187902 (2004).

\bibitem{Plenio} M. B. Plenio, and S. F. Huelga, New. J. Phys. \textbf{10}, 113019 (2008).

\bibitem{bosonsampling1} J. Sping \emph{et al.}, Science \textbf{339},  798-801 (2013).

\bibitem{bosonsampling2} M. Broome \emph{et al.}, Science \textbf{339},  6121 (2013).

\bibitem{bosonsampling3} M. Tillmann \emph{et al.}, Nature Photon \textbf{7}, 540 (2013). 

\bibitem{bosonsampling4} A. Crespi \emph{et al.}, Nature Photon. \textbf{7}, 545 (2013).

\bibitem{Grunbaum1} F. A. Grunbaum, L. Velazquez, A. H. Werner, R. F. Werner, Commun. Math. Phys. \textbf{320}, 543–569 (2013).

\bibitem{Grunbaum2} C. Cedzich, F. A. Grunbaum, C. Stahl, L. Velazquez, A. H. Werner and R. F. Werner, J. Phys. A: Math. Theor. \textbf{49},  21LT01 (2016).

\bibitem{validation1} J. Carolan \emph{et al.}, arXiv:1311.2913 (2013).

\bibitem{validation2} N. Spagnolo \emph{et al.}, arXiv:1311.1622 (2013).

\bibitem{Childs} A. M. Childs, Phys. Rev. Lett. \textbf{102}, 180501 (2009).

\bibitem{PuentesPRA} A. Aiello, G. Puentes, D. Voigt, J. P. Woerdman, Phys. Rev. A \textbf{75} (6), 062118 (2007).

\bibitem{PuentesOL} G. Puentes, D. Voigt, A. Aiello, J. P. Woerdman, Opt. Lett. \textbf{30} (23), 3216 (2005).
\bibitem{Peruzzo} A. Peruzzo \emph{et al.}, Science \textbf{329}, 1500-1503 (2010).

\bibitem{OBrien} K. Poulios \emph{et al.}, arXiv:1308.2554 (2013).

\bibitem{Silberhorn2D} A. Schreiber \emph{et al.}, Science \textbf{ 336}, pp. 55-58 (2012).

\bibitem{photons}A. Schreiber \emph{et al.}, Phys. Rev. Lett. \textbf{104}, 050502 (2010).

\bibitem{Silberhorn} A. Schreiber \emph{et al.}, Phys. Rev. Lett. \textbf{106}, 180403 (2011).

\bibitem{ValleyDirac1}A. Rycerz, J. Tworzydlo, and C. W. Beenakker, Nature Phys. \textbf{3}, 172 (2007). 

\bibitem{ValleyDirac2} D. Xiao, W. Yao, and Q. Niu, Phys. Rev. Lett.  \textbf{99}, 236809 (2007).

\bibitem{ValleyDirac3}D. Xiao, G-B. Liu, W. Feng, X. Xu, and W. Yao, Phys. Rev. Lett. \textbf{108}, 196802 (2012).

\bibitem{Gunawan}O. Gunawan, Y. P. Shkolnikov, K. Vakili, T. Gokmen, E. P. De Poortere, and M. Shayegan, Phys. Rev. Lett. \textbf{97}, 186404 (2006).

\bibitem{photons2} M. A. Broome \emph{et al.}, Phys. Rev. Lett. \textbf{104}, 153602 (2010).

\bibitem{ions}F. Zahringer \emph{et al.}, Phys. Rev. Lett. \textbf{104}, 100503 (2010).

\bibitem{coldatoms}M. Karski \emph{et al.}, Science \textbf{325}, 5947 (2009).

\bibitem{Pauli} M. Nielsen, and I. Chuang, \emph{Quantum Computation and Quantum Information}, Cambridge University Press (2000).
 
\bibitem{BrunPRL2003}T. A. Brun, H. A. Carteret, and A. Ambainis, Phys. Rev. Lett. \textbf{91}, 130602 (2003).
\bibitem{ZakDemler} M. Atala, M. Aidelsburger, J. Barreiro, D. Abanin, T. Kitagawa, E. Demler, I. Bloch, Nature Phys. \textbf{9}, 795 (2013).

\bibitem{ZakLonghi} S. Longhi, Opt. Lett. \textbf{38}, 3716 (2013).

\bibitem{HolonomicWhite} J. C. Loredo, M. A. Broome, D. H. Smith, and A. G. White, Phys. Rev. Lett. \textbf{112}, 143603 (2014).


\end{thebibliography}
\end{document}